# Exponential parameterization of neutrino mixing matrix with account for CP-violation data

K.Zhukovsky[1], F.M.Murilio[2].

**Abstract**

The exponential parameterization of Pontecorvo-Maki-Nakagawa-Sakata mixing matrix for neutrino is discussed. The exponential form allows easy factorization and separate analysis of the CP-violating and Majorana terms. Based upon the recent experimental data on the neutrino mixing, the values for the exponential parameterization matrix for neutrinos are determined. The matrix entries for the pure rotational part in charge of the mixing without CP-violation are derived. The complementarity hypothesis for quarks and neutrinos is demonstrated. The comparison of the results, based on most recent and on old data is held. The CP-violating parameter value is estimated, based on the so far imprecise experimental indications, regarding CP-violation for neutrinos. The unitarity of the exponential parameterisation and the CP-violating term transform is confirmed. The transform of the neutrino mass state vector by the exponential matrix with account for CP-violation is shown.

---

[1] *Deptartment of Theoretical Physics, Faculty of Physics, M.V.Lomonosov Moscow State University, Moscow 119991, Russia. Phone: +7(495)9393177 , e-mail: zhukovsk@physics.msu.ru*

[2] *Deptartment of Theoretical Physics, Faculty of Physics, M.V.Lomonosov Moscow State University, Moscow 119991, Russia. Phone: +7(495)9393177 , e-mail: francisco.melazzini@physics.msu.ru*

## 1. Introduction

One of the paramount achievements of Physics of 20th century was certainly the formulation of the Standard Model [1]–[3], which unifies the description of electromagnetic and weak interactions in one theory. Important role in the Standard Model is played by neutrinos. In the framework of the Standard Model neutrinos may have three flavours, matching three charged leptons, with which they interact by means of weak interaction. The proper states form full normalized orthogonal basis. Originally the Standard Model assumed massless neutrinos; later it was adopted to incorporate their mass. Existence of mass of neutrinos means the existence of at least three massive neutrino states $\nu_1$, $\nu_2$, $\nu_3$, and, also, it means the existence of the neutrino oscillations [4], i.e., neutrinos flavour change while they are propagating. Evidences for that were found in experiments for mixing of solar neutrinos [5], atmospheric neutrinos [6], and reactor neutrinos [7]. The phenomenon of neutrino mixing was predicted by Pontecorvo [8], [9]. The 2015 Nobel Prize in physics recognized the discovery of neutrino oscillations. The transforms from the mass state basis to the flavour state basis and vice versa is performed with the help of the Pontecorvo-Maki-Nakagawa-Sakata (PMNS) matrix. This transform can be considered similarly to the quark mixing by the CKM matrix. Familiar neutrino states $\nu_e$, $\nu_\mu$, $\nu_\tau$, are the linear combinations of neutrinos with different masses $\nu_1$, $\nu_2$, $\nu_3$:

$$\left|\nu_\alpha\right\rangle = \sum_{i=1,2,3} \mathbf{V}^*{}_{PMNS\,\alpha i}\left|\nu_i\right\rangle,\ \mathbf{V}_{PMNS\,\alpha i} \equiv \left\langle\nu_\alpha\middle|\nu_i\right\rangle, \tag{1}$$

where $\mathbf{V}_{PMNS}$ is the unitary PMNS mixing matrix [10]. Note the analogy with the mixing of the low elements of left components of quark spinors $\begin{pmatrix} u \\ d' \end{pmatrix}_L$, $\begin{pmatrix} c \\ s' \end{pmatrix}_L$, $\begin{pmatrix} t \\ b' \end{pmatrix}_L$. Lepton mixing presumes that a charged **W**-bozon can couple any mass state of charged leptons **(*e*, *μ*, *τ*)** with any mass state of neutrino. For example, $\mathbf{V}_{\alpha i}$ would be the amplitude of the bozon $\mathbf{W}^+$ decay in a pair of lepton type $\alpha$ and neutrino type $i$ and, hence, the production of the lepton $\alpha$ and neutrino state $\alpha$ implies that all neutrino mass states participate in it in a superposition. In what follows we do not consider sterile neutrino, [11], [12], [13], which does not interact with **W-** and **Z**-bozons. Thus, we end up with a unitary 3×3 mixing matrix **V**, factorized by the matrix:

$$\mathbf{V}_{PMNS} = \mathbf{V}\mathbf{P}_{Mjr}, \tag{2}$$

where

$$\mathbf{V} = \begin{pmatrix} c_{12}c_{13} & s_{12}c_{13} & s_{13}e^{-i\delta} \\ -s_{12}c_{23} - c_{12}s_{23}s_{13}e^{i\delta} & c_{12}c_{23} - s_{12}s_{23}s_{13}e^{i\delta} & s_{23}c_{13} \\ s_{12}s_{23} - c_{12}c_{23}s_{13}e^{i\delta} & -c_{12}s_{23} - s_{12}c_{23}s_{13}e^{i\delta} & c_{23}c_{13} \end{pmatrix}, \tag{3}$$

$$\mathbf{P}_{Mjr} = diag\left(e^{i\alpha_1/2}, e^{i\alpha_2/2}, 1\right), \tag{4}$$

$c_{ij} = \cos\theta_{ij}$, $s_{ij} = \sin\theta_{ij}$, $i,j$=1,2,3, and $\mathbf{P}_{Mjr} = diag\left(e^{i\alpha_1/2}, e^{i\alpha_2/2}, 1\right)$ describes possible Majorana nature of neutrinos by the phases $\alpha_1$ and $\alpha_2$. If $\alpha_{1,2} \neq 0$, then neutrinos are the Majorana particles, i.e. they are identical to their antiparticles; the phases $\alpha_1$, $\alpha_2$ play role in the processes, which do not preserve the lepton number. The role of the matrix **V** in the parameterization (3) is similar to that the CKM matrix plays in quark mixing [14], [16]–[19]. PMNS matrix is fully determined by four parameters: three mixing angles $\theta_{12}$, $\theta_{23}$, $\theta_{13}$ and the phase $\delta$, in charge of the CP-violation description [14]. Experimental values of the mixing angles are relatively well determined [14], [15], [20]–[22], [23]:

$$\theta_{12} \cong 33.36° \pm 0.8°, \tag{5}$$

$$\theta_{23} \cong 40.0° \pm 2°, \tag{6}$$

$$\theta_{13} \cong 8.66° \pm 0.45°. \tag{7}$$

Contrary to quark mixing angles, these are not small and the expansion in series of the only small parameter is not possible. Thus, there is no small parameter, like $\lambda=\sin\theta_{\text{Cabibbo}}\approx0.22$ [24], for neutrino mixing. Experimentally determined absolute values for the elements of PMNS-matrix read as follows [14]:

$$|\mathbf{V}| = \begin{pmatrix} 0.82 & 0.54 & 0.15 \\ 0.35 & 0.70 & 0.62 \\ 0.44 & 0.45 & 0.77 \end{pmatrix}. \tag{8}$$

Moreover, there are indications, that the CP-violating phase may have non-zero value; moreover, very approximately it is supposed to be as big as $\delta \approx 300°$ (see, [25], [26]).

## 2. Exponential mixing matrix

Exponential parameterization for neutrino mixing matrix was outlined in [27] and then in [28]; it is constructed similarly to that for quarks [29], [30]. Unitarity of the exponential mixing matrix $\mathbf{V} = \exp \mathbf{A}$ is guaranteed by the anti-Hermitian form of the exponent $\mathbf{A} = \begin{pmatrix} 0 & \lambda_1 & \lambda_3 e^{i\delta_{CP}} \\ -\lambda_1 & 0 & -\lambda_2 \\ -\lambda_3 e^{-i\delta_{CP}} & \lambda_2 & 0 \end{pmatrix}$ (see [32]), which depends on the mixing parameters $\lambda_i$, and on the CP-violation phase $\delta_{\mathrm{CP}}$. Note, that for $\delta_{\mathrm{CP}}=2\pi n$ we obtain simply a rotation matrix around the axis in space [30]. The matrix also becomes Real for $\delta_{\mathrm{CP}}=\pi(2n+1)$. The most important advantage of the exponential parameterization for the mixing matrix with respect to the commonly known standard parameterization is that the exponential parameterization allows easy separation of the contributions of the rotation part, the CP-violation and possible other terms in stand alone factors. This separation can be made in a variety of modes, which details we omit here; proper discussion was made, for example, in [30], [31]. The following unitary parameterization was proposed in [28]:

$$\widetilde{\mathbf{V}} = \mathbf{P}_{\mathbf{Rot}} \mathbf{P}_{\mathbf{CP}} \mathbf{P}_{\mathbf{Mjr}}, \tag{9}$$

where the rotation part is given by the Real exponential matrix

$$\mathbf{P}_{\mathbf{Rot}} = e^{\mathbf{A}_{Rot}} = \exp \begin{pmatrix} 0 & \lambda & \mu \\ -\lambda & 0 & -\nu \\ -\mu & \nu & 0 \end{pmatrix}, \tag{10}$$

the CP-violation is accounted for by

$$\mathbf{P}_{CP} = e^{\mathbf{A}_{CP}}, \tag{11}$$

and it contains imaginary component as follows:

$$\mathbf{A}_{\mathbf{CP}} = \begin{pmatrix} 0 & 0 & \mu\left(-1 + e^{i\delta_{CP}}\right) \\ 0 & 0 & 0 \\ \mu\left(1 - e^{-i\delta_{CP}}\right) & 0 & 0 \end{pmatrix}, \tag{12}$$

and the Majorana part

$$\mathbf{P}_{Mjr} = e^{\mathbf{A}_{Mjr}}, \tag{13}$$

depends on the Majorana phases in the exponential:

$$\mathbf{A}_{\mathbf{Mjr}} = i \begin{pmatrix} \alpha_1/2 & 0 & 0 \\ 0 & \alpha_2/2 & 0 \\ 0 & 0 & 0 \end{pmatrix}. \tag{14}$$

The details of the splitting between CP-conserving and CP-violating terms in the above parameterization can be found in [28] (also compare with [30], [31]). The values of the Majorana phases $\alpha_1$ and $\alpha_2$ are at present undetermined; the value of the CP-violating phase $\delta_{CP}$ can be figured out from the existing experimental indications and estimations (see, for example, [25], [26]).

The rotation matrix can be conveniently presented in the form of the rotation in the angle $\Phi$ around the axis, given by the vector $\vec{\mathbf{n}} = (n_x, n_y, n_z)$:

$$\mathbf{M}(\vec{\mathbf{n}}, \Phi) = e^{\Phi \mathbf{N}} = \begin{pmatrix} M_{xx} & M_{xy} & M_{xz} \\ M_{yx} & M_{yy} & M_{yz} \\ M_{zx} & M_{zy} & M_{zz} \end{pmatrix}, \ \mathbf{N} = \begin{pmatrix} 0 & -n_z & n_y \\ n_z & 0 & -n_x \\ -n_y & n_x & 0 \end{pmatrix}. \qquad (15)$$

In this form it represents three-dimensional rotation generator. Then (15) links the entries of the rotational matrix (10) in the exponential parameterization (9) with the rotation angle

$$\Phi = \pm\sqrt{\lambda^2 + \mu^2 + \nu^2} \qquad (16)$$

around the axis $\vec{\mathbf{n}} = (n_x, n_y, n_z)$ with the following coordinates:

$$n_x = \frac{\nu}{\Phi}, n_y = \frac{\mu}{\Phi}, \ n_z = -\frac{\lambda}{\Phi}. \qquad (17)$$

Thus, formulae (17)–(16) relate the elements of the exponential parameterization with the single axis-angle rotation matrix components. Note, that the angles of the rotation in the standard parameterization matrix (3) $c_{ij} = \cos\theta_{ij}$ and $s_{ij} = \sin\theta_{ij}$ are different from those in our parameterization (9). Omitting the Majorana part, the exponential parameterization (9) reads as follows:

$$\tilde{\mathbf{V}} = \mathbf{M}\mathbf{P}_{\mathbf{CP}} = \begin{pmatrix} M_{xx}\cos 2\Omega + \omega_- M_{xz}\sin 2\Omega & M_{xy} & M_{xz}\cos 2\Omega + \omega_+ M_{xx}\sin 2\Omega \\ M_{yx}\cos 2\Omega + \omega_- M_{yz}\sin 2\Omega & M_{yy} & M_{yz}\cos 2\Omega + \omega_+ M_{yx}\sin 2\Omega \\ M_{zx}\cos 2\Omega + \omega_- M_{zz}\sin 2\Omega & M_{zy} & M_{zz}\cos 2\Omega + \omega_+ M_{zx}\sin 2\Omega \end{pmatrix}, \qquad (18)$$

where

$$\Omega = \mu \sin\frac{\delta_{CP}}{2}, \ \omega_\pm = e^{\frac{i}{2}(\pi \pm \delta_{CP})}. \qquad (19)$$

The values of the entries of the rotation matrix $M_{ij}$ can be derived from the following tensor identity (see [33]):

$$M_{ij} = (1 - \cos\Phi) n_i n_j + \delta_{ij} \cos\Phi - \varepsilon_{ijk} n_k \sin\Phi \,, \; i, j, k = x, y, z \,, \tag{20}$$

where we denote $\delta_{ij}$ the Kronecker symbol, $\varepsilon_{ijk}$ is the Levi-Civita symbol, $n_i$ are the components of the vector $\vec{\mathbf{n}} = (n_x, n_y, n_z)$ and $\Phi$ is the rotation angle. The expressions, relating the entries of the standard parameterization $c_{ij}$ and $s_{ij}$ with $\vec{\mathbf{n}}$ and $\Phi$, can be derived from (18), (19), (20) and (3), but they are very cumbersome and we omit them for brevity. From the experimental data [25] by using matrix equations and expansions we obtain for the rotation vector in 3D space the following coordinates:

$$n_x \cong 0.702, \; n_y \cong 0.394, \; n_z \cong 0.593 \,, \tag{21}$$

and for the rotation angle around this axis we obtain

$$\Phi \cong 49.8^\circ \,. \tag{22}$$

The precision of the above values is determined by the errors in the experimental data evaluation and is about of 4%. From (15)–(17) we obtain the following values for the entries of the rotation matrix (15) in the exponential parameterization:

$$\mathbf{A}_{Rot} \cong \begin{pmatrix} 0 & 0.516 & -0.342 \\ -0.516 & 0 & 0.611 \\ 0.342 & -0.611 & 0 \end{pmatrix}, \quad \begin{matrix} \lambda \cong 0.516, \\ \mu \cong -0.342, \\ \nu \cong -0.611. \end{matrix} \tag{23}$$

This result is in agreement with the experimental data and with respective standard parameterization in the Pontecorvo-Maki-Nakagawa-Sakata matrix (2). Moreover, based upon the values of the mixing angles for quarks $\theta_{Q_{12}} = 13.14^\circ$, $\theta_{Q_{23}} = 2.43^\circ$, $\theta_{Q_{13}} = 0.23^\circ$, we determine the direction of the rotation vector in space (17) for the exponential parameterization for quarks as follows:

$$\vec{\mathbf{n}}_{quark} = (0.1829, 0.0206, 0.9831) \,. \tag{24}$$

Now, upon the comparison with the above determined coordinates of the rotation vector for neutrinos

$$\vec{\mathbf{n}}_{neutrino} = (0.7021, 0.3936, 0.5934) \,, \tag{25}$$

we note that $\vec{\mathbf{n}}_{quark}$ and $\vec{\mathbf{n}}_{neutrino}$ constitute the angle of ≈44°, as demonstrated in Fig. 1.

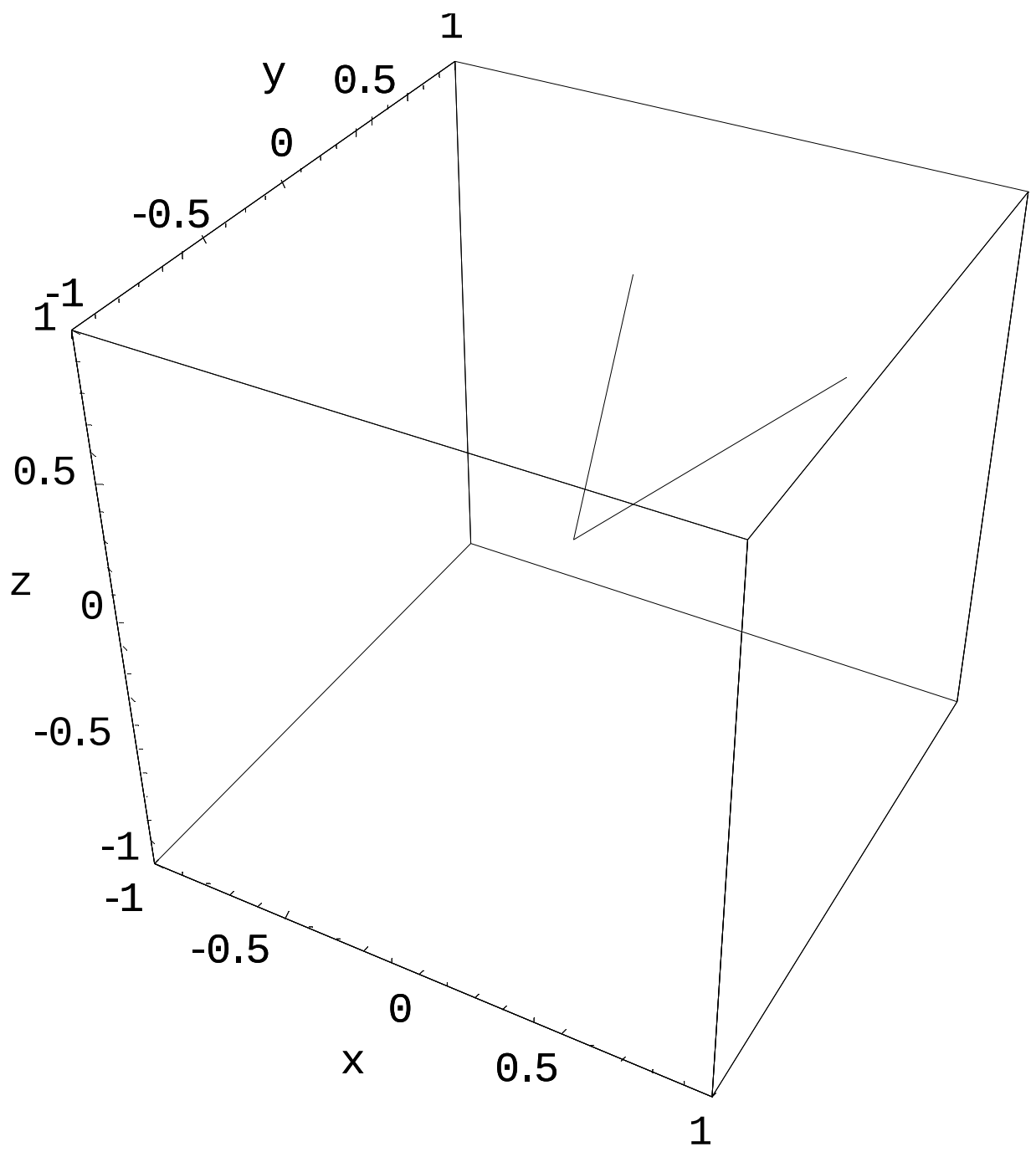

**Fig. 1 The axes of rotation for quarks and neutrinos form the angle of ≈45°.**

This fact is interesting itself and it is the demonstration of the so-called hypothesis of complementarity for neutrinos and quarks [34], [35], according to which, the rotation axes for quarks and neutrinos form the 45° angle; however, this last statement is rather an observation since it does not have solid theoretical fundaments and physical reasons. Note, that the obtained value of 44° differs from 45° by ≈2%, which is within the margin of errors of the original experimental data sets, which determines the entries of the exponential mixing matrix and the rotation vectors directions.

## 3. Exponential parameterization and CP-violation.

Now let us take advantage of the possibility, given by the exponential parameterization, which allows us to factorize separately the contributions of the rotation, the CP-violation and the Majorana term. We can write the matrix product $\mathbf{P}_{\mathbf{CP}}\mathbf{P}_{\mathbf{Mjr}}$ in the following form, which reminds the rotation in the angle 2Ω with the proper weights for each entry:

$$\mathbf{V_{MCP}} = \mathbf{P_{CP}}\mathbf{P_{Mjr}} = \begin{pmatrix} e^{i\frac{\alpha_1}{2}}\cos 2\Omega & 0 & \omega_+ \sin 2\Omega \\ 0 & e^{i\frac{\alpha_2}{2}} & 0 \\ \omega_ e^{i\frac{\alpha_1}{2}} \sin 2\Omega & 0 & \cos 2\Omega \end{pmatrix} =$$

$$\begin{pmatrix} e^{i\frac{\alpha_1}{2}}\cos\left(2\mu\sin\frac{\delta_{CP}}{2}\right) & 0 & e^{\frac{i}{2}(\pi+\delta_{CP})}\sin\left(2\mu\sin\frac{\delta_{CP}}{2}\right) \\ 0 & e^{i\frac{\alpha_2}{2}} & 0 \\ e^{\frac{i}{2}(\pi-\delta_{CP}+\alpha_1)}\sin\left(2\mu\sin\frac{\delta_{CP}}{2}\right) & 0 & \cos\left(2\mu\sin\frac{\delta_{CP}}{2}\right) \end{pmatrix}. \tag{26}$$

Omitting the Majorana part, the complexity due to the CP-violation vanishes if either $\Omega = \pm\pi n/2$, or if $\omega_\pm \in \text{Reals}$. The value of $\Omega$ is limited by $\mu$, which is small, and the above expression (26) becomes Real in the case of $\Omega = 0$, which, moreover, returns the unity matrix $\mathbf{P_{CP}} = \mathbf{I}$ for $\delta_{CP} = \pm 2\pi n$. The absolute values of the entries of the CP-violating matrix are at their maximum for $\delta_{CP} = \pi \pm 2\pi n$, but then the whole $\mathbf{P_{CP}}$ matrix becomes Real and the CP violation in fact vanishes. The major contribution of the complex term due to the CP-violation is achieved for $\delta_{CP} = \pi/2 \pm \pi n$. All of the above said is in complete analogy with the behaviour of the CP-violating phase $\delta$ in the standard parameterization. Based upon the present knowledge of the CP-violation for neutrinos and on experimental indications and estimations [25], [26] and accounting for the above discussion, the CP-violating phase in the exponential parameterisation has the value $\delta_{CP} \cong -60°$ identical to $\delta \approx 300°$. Then we obtain

$$\omega_\pm\Big|_{\delta_{CP}=-60°} e^{\frac{i\pi}{2}\left(1\mp\frac{1}{3}\right)} \cong \pm 0.5 + 0.866\,i\ . \tag{27}$$

Note, that the accuracy of the existing indications with regard to the CP-violation is rather low ($+60-120\%$) and thus the above values are also approximate. The parameter $\Omega$ is not very sensitive to the exact value of the angle $\delta_{CP}$ and it's absolute value varies from 0 to the maximum of $|\mu| = 0.342$ for $\delta_{CP} = \pi \pm 2\pi n$. Note, that in any case the value of $\Omega$ is small and $\cos\Omega \cong 1$; For $\delta_{CP} = -60°$ we obtain

$$\Omega = \mu \sin\frac{\delta_{CP}}{2} = \mu \sin(-30°) = -\mu/2 = 0.171,$$
$$\cos 2\Omega = \cos\mu \cong 0.942, \quad \sin 2\Omega = -\sin\mu \cong 0.335. \tag{28}$$

Then with account for Majorana phases and for $\delta_{CP} \cong -60°$, we obtain the following matrix:

$$\mathbf{V_{MCP}} = \mathbf{P_{CP}}\mathbf{P_{Mjr}} = \begin{pmatrix} e^{i\frac{\alpha_1}{2}}0.942 & 0 & 0.335\omega_+ \\ 0 & e^{i\frac{\alpha_2}{2}} & 0 \\ 0.335e^{i\frac{\alpha_1}{2}}\omega_- & 0 & 0.942 \end{pmatrix} \cong$$
$$\begin{pmatrix} 0.942e^{i\frac{\alpha_1}{2}} & 0 & 0.168 + 0.290i \\ 0 & e^{i\frac{\alpha_2}{2}} & 0 \\ e^{i\frac{\alpha_1}{2}}(-0.168 + 0.290i) & 0 & 0.942 \end{pmatrix}. \tag{29}$$

For Dirac neutrinos at the extremities of the range of the CP-violating phase for $\delta_{CP} = 0°$ we get the unity matrix $\mathbf{P_{CP}} = \mathbf{I}$, while for $\delta_{CP} \cong -180°$: $\omega_\pm\big|_{\delta_{CP}=-180°} = e^{\frac{i\pi}{2}(1\mp1)} = \pm 1$ and the Real mixing term

$$\mathbf{P_{CP}} \equiv \mathbf{V_{MCP}} = \begin{pmatrix} 0.775 & 0 & 0.632 \\ 0 & 1 & 0 \\ -0.632 & 0 & 0.775 \end{pmatrix}. \tag{30}$$

The Majorana phases interplay with the CP-phase only in the entry (3,1) in the factor $e^{i\frac{\alpha_1}{2}}\omega_-$. Otherwise, the Majorana phases just bring more complexity in the result. For non-Majorana, but Dirac particles, the above obtained matrix (29) represents just a slight deviation from the unitary matrix **I**:

$$\mathbf{P_{CP}} \equiv \mathbf{V_{MCP}} \cong \begin{pmatrix} 1 & 0 & 0.168 + 0.290i \\ 0 & 1 & 0 \\ -0.168 + 0.290i & 0 & 1 \end{pmatrix}. \tag{31}$$

It is now evident that for non-Majorana particles the CP-violating term (29) is the mixing matrix for two lepton generations: electron, taon and proper neutrinos. Employing the generating functions for the Bessel functions $\cos(x\sin\alpha)=\sum_{n=-\infty}^{\infty}J_n(x)\cos n\alpha$ and $\sin(x\sin\alpha)=\sum_{n=-\infty}^{\infty}J_n(x)\sin n\alpha$, we easily derive the following expression for the general form of the matrix $\mathbf{V_{MCP}}$, in which the contributions of the CP-violating phase $\delta_{CP}$ and of the rotation matrix parameter $\mu$ are separated:

$$\mathbf{V_{MCP}}=\begin{pmatrix} e^{i\frac{\alpha_1}{2}}\sum_{n=-\infty}^{\infty}J_n(2\mu)\cos\left(\frac{n\delta_{CP}}{2}\right) & 0 & \omega_+\sum_{n=-\infty}^{\infty}J_n(2\mu)\sin\left(\frac{n\delta_{CP}}{2}\right) \\ 0 & e^{i\frac{\alpha_2}{2}} & 0 \\ \omega_e^{i\frac{\alpha_1}{2}}\sum_{n=-\infty}^{\infty}J_n(2\mu)\sin\left(\frac{n\delta_{CP}}{2}\right) & 0 & \sum_{n=-\infty}^{\infty}J_n(2\mu)\cos\left(\frac{n\delta_{CP}}{2}\right) \end{pmatrix}. \quad (32)$$

The above result (32) is valid for arbitrary values of $\delta_{CP}$, $\alpha_1$, $\alpha_2$. For $\alpha_1$=0 (32) it reduces to symmetric form. With account for the obtained values of the entries of the exponential parameterization of the CP-violating term and for non-zero Majorana phases $\alpha_1$, $\alpha_2$, the vector of the mixed neutrino state is transformed by $\mathbf{V^*_{MCP}}$ as follows:

$$|\nu_\alpha\rangle=\begin{pmatrix} 0.942e^{-\frac{i}{2}\alpha_1}|\nu_1\rangle+0.335e^{-\frac{\pi}{3}i}|\nu_3\rangle \\ e^{-i\frac{\alpha_2}{2}}|\nu_2\rangle \\ 0.335e^{-i\left(\frac{\alpha_1}{2}+\frac{2\pi}{3}\right)}|\nu_1\rangle+0.942|\nu_3\rangle \end{pmatrix}\cong\begin{pmatrix} |\nu_1\rangle+(0.168-0.290\,i)|\nu_3\rangle \\ e^{-i\frac{\alpha_2}{2}}|\nu_2\rangle \\ -(0.168+0.290\,i)e^{-\frac{i}{2}\alpha_1}|\nu_1\rangle+|\nu_3\rangle \end{pmatrix}. \quad (33)$$

We underline that the above transform by $\mathbf{V_{MCP}}$ as well as the transform by purely rotational part $\mathbf{P_{Rot}}$, is unitary. It can be verified directly with the help of the Hermite-conjugated matrix:

$$\mathbf{V}_{\mathbf{MCP}}^{-1}\cdot\mathbf{V}_{\mathbf{MCP}}=\mathbf{V}_{\mathbf{MCP}}^{+}\cdot\mathbf{V}_{\mathbf{MCP}}=\mathbf{I}. \quad (34)$$

This ensures the unitarity of the whole exponential parameterization (9) of the Pontecorvo-Maki-Nakagawa-Sakata mixing matrix:

$$\tilde{\mathbf{V}}^{-1}\cdot\tilde{\mathbf{V}}=\tilde{\mathbf{V}}^{+}\cdot\tilde{\mathbf{V}}=\mathbf{I}. \quad (35)$$

## Conclusions

The exponential parameterization of the mixing matrix for neutrinos is explored with account for the present experimental data. The proper entries of the exponential mixing matrix are determined; the CP-violating term in the exponential parameterization is estimated. Based upon the accuracy of the experimental data, the range of the values for the parameters of the neutrino mixing matrix is given. Without CP-violation the neutrino mixing represents in fact the geometric rotation in three-dimensional space. In this simple case mixing can be viewed as the rotation in the angle Φ around the axis in three-dimensional space. This interpretation follows straight from the structure of the exponential mixing matrix. Evidently, there is no mixing for Φ=0, when the mixing matrix without CP-violation reduces to the unit matrix **I**. Based upon the recent data, we have obtained the value for the rotation angle $\Phi \cong 49.8°$ and the coordinates of the rotation vector $\vec{\mathbf{n}} = (0.702, 0.394, 0.593)$. This value of the rotation angle is some smaller than that, based upon the tribimaximal parameterization: $\Phi_{TBM} \cong 56.6°$. Moreover, the direction of the rotation vector differs form that of the vector for the TBM matrix: $\vec{\mathbf{n}}_{TBM} = (0.7858, 0.2235, 0.5777)$. The difference in their directions in 3D space is 11°. Interestingly, the angle between the axes of rotation for quarks and neutrinos remains unchanged and equals ≈45°, despite the change of 11° in the direction of the neutrinos rotation axis, verified in the last 10 years. This demonstrates the hypothesis of complementarity for quarks and neutrinos [34], [35].

The exponential parameterization allows factorization of the CP and the Majorana contributions and evidences that the CP-term can also be viewed as a sort of rotation with different weights for the matrix entries. We have calculated the entries of the rotational mixing matrix $\lambda \cong 0.516$, $\mu \cong -0.342$, $\nu \cong -0.611$ (see (23)) and we have estimated the entries of the CP-violating matrix in the exponential parameterization (see (29), (31)) from the current indications on CP-violation: $\delta_{CP} \sim -60°$. This value is quite approximate due to uncertain experimental data, regarding CP-violation for neutrinos. We calculated the CP-violating exponential matrix for the extremities of the range of $\delta_{CP}$ from 0° to 180°. The rotation angle 2Ω for the CP-violating matrix is rather small: $\Omega = \mu/2$. By means of the exponential parameterization one can easily transform the neutrino state vector, distinguishing the CP-violating terms for each type of neutrino. In the case of Dirac neutrinos, if $\delta_{CP} = \pm 2\pi n$, we get pure rotation, since $\mathbf{P_{CP}} = \mathbf{I}$. If $\delta_{CP} = \pi \pm 2\pi n$, then we end with the Real $\mathbf{P_{CP}}$ matrix for the CP-violating term, which means the absence of the CP-violation in this case. For Majorana neutrinos the mass state vector $\mathbf{v}_1$, $\mathbf{v}_2$, $\mathbf{v}_3$ is transformed with the complex weights, as demonstrated in (33).

Exponential presentation of the mixing matrix and obtained with its help results and interpretations can be useful for treatment and analysis of new experimental data, regarding the neutrino oscillations in currently running experiments as well as in planned experimental projects.

Acknowledgements: We would like to thank professor A.V. Borisov from the Moscow State University, Russia, for his useful suggestions, fruitful discussions and important advises.

## References:

[1]. S.Weinberg,Phys. Rev. Lett, v.19, 1264, 1967.

[2]. A.Salam, Elementary Particle Theory, Ed. by N.Svartholm, Almquist Forlag AB, 1968.

[3]. S.L.Glashow, Nucl. Phys., v.22,.579, 1961.

[4]. Barger, Vernon; Marfatia, Danny; Whisnant, Kerry Lewis (2012). *The Physics of Neutrinos*. Princeton University Press.

[5]. T.Ataki et al. (KamLAND Collaboration), Phys. Rev. Lett. 94, 081801 (2005)

[6]. Y.Ashie et al. (Super-Kamiokande Collaboration), Phys. Rev. D 71, 112005 (2005).

[7]. M.Apollonio et.al. (CHOOZ Collaboration), Eur. Phys. J. C 27, 331 (2003).

[8]. B.Pontecorvo, Zh.Eksp.Teor.Fiz., vol.33, p.549. (1957).

[9]. B.Pontecorvo, Zh.Eksp.Teor.Fiz., vol.53, p.1717. (1967) [Sov.Phys. JETP vol.26, 984 (1968)].

[10]. Z.Maki, M.Nakagawa and S.Sakata, Prog. Teor. Phys. 28, 870 (1962).

[11]. H.Murayama and T.Yanagida, Phys.lett. B 520, 263 (2001).

[12]. G.Barenboim et.al., JHEP 0210, 001 (2002).

[13]. M.C.Gonzalez-Garcia, M.Maltoni and T.Schwetz, Phys.Rev. D 68, 053007 (2003).

[14]. K.A. Olive et al. (Particle Data Group), Chin. Phys. C, 38, 090001 (2014).

[15]. H. Minako, K. Yee, O. Naotoshi, T. Tatsu, A Simple Parameterization of Matter Effects on Neutrino Oscillations;YITP vol.05-52; p.73P(2006), arXiv:hep-ph/0602115v1.
[16]. L.-L. Chau and W.-Y. Keung, Phys. Rev. Lett. 53, 1802 (1984).
[17]. H. Harari and M Leurer, Phys. Lett. B181, 123 (1986).
[18]. H. Fritzsch and J. Plankl, Phys. Rev. D 35, 1732 (1987).
[19]. F.J. Botella and L.-L. Chao, Phys . Lett. B 168, 97 (1986).
[20]. B.Aharmin *et.al* (The SNO Collaboration), nucl-ex/0502021.
[21]. T.Araki *et.al* (The KamLAND Collaboration ) Phys.Rev.Lett. 94, 081801 (2005).
[22]. Y.Suzuki, presented at the XXII Int Symp. On Lepton and Photon Interactions at High Energies (Lepton-Photon 2005), Uppsala, Sweden, July 2005.
[23]. C. Bemporad (The Chooz Collaboration), Nucl. Phys. Proc. Suppl. **77** (1999) 159.
[24]. L.Wolfenstein, Phys.Rev.lett., 51 (1983) 1945.
[25]. Stephen F. King, arXiv:1510.02091v1 [hep-ph] 2015.
[26]. Ernest Ma, arXiv:1510.02501v1 [hep-ph] 2015.
[27]. A. Strumia, F. Vissani, "Neutrino masses and mixings and..." hep-ph/0606054 (2006).
[28]. G Datolli, K. V. Zhukovsky. Eur. Phys. J. C 55, 547–552 (2008).
[29]. G.Dattoli, K.Zhukovsky, Eur. Phys. J. C 50, 817-821 (2007).
[30]. G.Dattoli, K.Zhukovsky, Eur. Phys. J. C 52, N3, 591-595 (2007).
[31]. G. Dattoli, K.V. Zhukovsky, Physics of Atomic Nuclei, 71, N10, 1807-1812 (2008).
[32]. G.Dattoli, E.Sabia, A.Torre, Nuovo Cimento A, 109A (1996) 1425.
[33]. H.Goldstein, Classical Mechanics. Cambridge, MA: Addison-Wesley, 1950 399p.
[34]. H.Minakata, A.Yu.Smirnov, Phys.Rev. D70, 073009 (2004).
[35]. M.Raidal, Phys. Rev. Lett. 93, 161801 (2004).